\documentclass[10pt,conference,compsocconf]{IEEEtran}
\usepackage{times}
\usepackage{caption}
\usepackage{color}
\IEEEoverridecommandlockouts

\ifCLASSINFOpdf
   \usepackage[pdftex]{graphicx}
\else
\fi
\usepackage[cmex10]{amsmath}
\DeclareMathOperator*{\argmax}{arg\,max}

\begin{document}
\pagenumbering{gobble}
%
\title{\textbf{\Large An Agent-Based Model of Delegation Relationships With Hidden-Action: On the Effects of Heterogeneous Memory on Performance}\\[0.2ex]}

\author{\IEEEauthorblockN{~\\[-0.4ex]\large Patrick Reinwald\\[0.3ex]\normalsize}
\IEEEauthorblockA{Department of Management Control  \\and Strategic Management\\
University of Klagenfurt\\
Klagenfurt, Austria\\
Email: {patrick.reinwald@aau.at}\\
ORCID: 0000-0002-2907-7939}
\and
\IEEEauthorblockN{~\\[-0.4ex]\large Stephan Leitner\\[0.3ex]\normalsize}
\IEEEauthorblockA{Department of Management Control  \\and Strategic Management\\
	University of Klagenfurt\\
	Klagenfurt, Austria\\
	Email: {stephan.leitner@aau.at}\\
ORCID: {0000-0001-6790-4651}}
\and
\IEEEauthorblockN{~\\[-0.4ex]\large Friederike Wall\\[0.3ex]\normalsize}
\IEEEauthorblockA{Department of Management Control  \\and Strategic Management\\
	University of Klagenfurt\\
	Klagenfurt, Austria\\
	Email: {friederike.wall@aau.at}\\
ORCID: {0000-0001-8001-8558}}
}

%


\maketitle

\begin{abstract}

We introduce an agent-based model of delegation relationships between a principal and an agent, which is based on the standard-hidden action model introduced by Holmstr\"om and, by doing so, provide a model which can be used to further explore theoretical topics in managerial economics, such as the efficiency of incentive mechanisms. We employ the concept of agentization, i.e., we systematically transform the standard hidden-action model into an agent-based model. Our modeling approach allows for a relaxation of some of the rather "heroic" assumptions included in the standard hidden-action model, whereby we particularly focus on assumptions related to the (i) availability of information about the environment and the (ii) principal's and agent's cognitive capabilities (with a particular focus on their learning capabilities and their memory). 
Our analysis focuses on how close and how fast the incentive scheme, which endogenously emerges from the agent-based model, converges to the solution proposed by the standard hidden-action model. Also, we investigate whether a stable solution can emerge from the agent-based model variant. 
The results show that in stable environments the emergent result can nearly reach the solution proposed by the standard hidden-action model. Surprisingly, the results indicate that turbulence in the environment leads to stability in earlier time periods.
\end{abstract}




%
\IEEEpeerreviewmaketitle

\section{Introduction}

The standard hidden-action model introduced by Holmstr\"om \cite{Holmstrom.1979} describes, in general, a delegation relation between a principal and an agent. It covers a situation in which the principal delegates a task to an agent. The agent selects an effort level in order to carry out this task, which is not observable by the principal (it is \textit{hidden}). The agent's effort together with some environmental impact produces outcome which is to be shared between the principal and the agent. The principal's objective is to maximize her share of the outcome associated with the task while the agent strives for maximizing his share of the outcome minus his disutility from making effort to carry out the task. The standard hidden-action model proposes an incentive scheme (i.e., a rule to share the outcome) which aligns the agent's and the principal's objective so that the principal's utility finds its maximum. As only the outcome (but not the effort level) is observable for the principal, the sharing rule is based on outcome only. 

In order to derive the optimal sharing rule, agency theory makes rather "heroic'" assumptions about the capabilities of both parties with respect to (i) information processing capacity, (ii) availability of information, and (iii) capability to find the optimal solution immediately. Axtell \cite{Axtell.2007} argues that the most "heroic" assumptions are agent homogeneity, non-interactiveness, and the existence of equilibrium solutions and refers to them as "neoclassical sweetspot". 
The result of these rather "heroic" assumptions is that the explanatory and the predictive power regarding real world problems substantially decreases \cite{Schneeweiss.1987}.
Critics often refer to principal-agent models as "toy problems" and argue that solutions derived from such "toy problems", where authors can "assume complicating things away", are of limited use and only merely help to solve real world problems \cite{Jensen.1983}-\nocite{Eisenhardt.1989}\cite{Shapiro.2005}. 

We take up on this critique and put the "heroic" assumptions included in the standard hidden-action model \cite{Holmstrom.1979} in the focus of this paper. In the vein of Leitner and Wall \cite{Leitner.2020}, we transfer the standard hidden-action model into an agent-based model following a procedure introduced by Guerrero and Axtell \cite{Guerrero.2011} and Leitner and Behrens \cite{Leitner.2015}, which allows us to relax some of the included assumptions. In particular, we relax the assumptions related to the principal's and the agent's (i) information processing capacity and (ii) the availability of information. We model situations in which the principal and the agent no longer have full information about the environment but have to learn this information over time. In consequence, they can no longer find the optimal (second-best) solution immediately, like it is the case in the standard hidden-action model. We, therefore, endow the principal and the agent with the capability to search for the best possible incentive scheme over time by employing a hillclimbing algorithm. 
 Please notice that the optimal solution, which is derived from the standard hidden-action model for cases in which the the agent's effort is \textit{not} observable, is referred to as \textit{second-best solution}. The first-best solution, on the contrary, assumes that the agent's effort is observable \cite{Lambert.2001}, which, in consequence, means that no incentive problem in the above outlined sense arises. 
 
The remainder of this paper is organized as follows: Section \ref{sec2} introduces two variants of the hidden-action model. First the main features of the standard hidden-action model are presented. We, then introduce the agent-based model variant which relaxes some of the assumptions included in the standard hidden-action model. In Section \ref{sec3}, we elaborate on the simulation setup, and introduce and discuss the results. Section \ref{sec4} concludes the paper and gives an outlook on future work.

\section{The hidden-action Model}
\label{sec2}
This section summarizes the main features of the standard hidden-action model introduced in \cite{Holmstrom.1979} and proposes an agent-based representation of the hidden-action problem in which the principal and the agent are endowed with limited and heterogeneous memory. 

\subsection{The standard hidden-action model}
The standard hidden-action model, which describes a delegation relation between a principal and an agent,  was first described by Holmstr\"om  \cite{Holmstrom.1979}. This model covers a situation in  which a principal offers an agent a contract upon a task to be carried out and a sharing rule over the generated outcome, among other things. If the agent agrees on the condition stated in the contract, he exerts effort to complete the specified task. Together with an exogenous factor, his effort generates outcome, but this also leads to disutility for him. For the principal, however, the effort the agent had carried out is unobservable, which results in a situations where only the outcome can be used as a basis for the sharing rule.
Both the principal and the agent are individual utility maximizers \cite{Lambert.2001,Eisenhardt.1989}. Furthermore, it is assumed that the principal is risk neutral and characterized by her utility function
\begin{equation}
\label{utilityP}
U_P(x,s) = x-s(x)~,
\end{equation} whereby $ x $ represents the generated outcome and $ s =s(x) $ is the function for the sharing rule. As mentioned before, the outcome
\begin{equation}
\label{prodctionfunction}
x=f(a,\theta)~,
\end{equation}
  is a function of the agent's effort $ a $ and the exogenous factor $ \theta $. The agent, who is assumed to be risk averse, is characterized by the utility function
  \begin{equation}
   U_A(s,a)=V(s)-G(a)~,
  \end{equation}
    where $ V(s) $ represents the generated utility from the compensation and $ G(a) $ denotes to disutulity from the exerted effort. For the contract to be effective, two additional constraints have to be fulfilled. First, the participation constraint 
    \begin{equation}
    E(U_a(s,a)) \geq \bar{U}~,
    \end{equation}
    which assures that the agent gets at least the utility he would get from the best outside option $ \bar{U} $. Second, the incentive compatibility constraint 
    \begin{equation}
     a \in \argmax_{a} \: E\{U_A(s(x),a')\}~,
    \end{equation}
     which aligns the agent's goal (maximizing his utility) with the goal of the principal. For an extensive review and the formal solution to this problem, the reader is, for example, referred to Holmstr\"om \cite{Holmstrom.1979} and Lambert \cite{Lambert.2001}.
     
\subsection{Agent-based model variant} 

\begin{figure}
	\label{flow}
	\centering
	\includegraphics[scale=0.32]{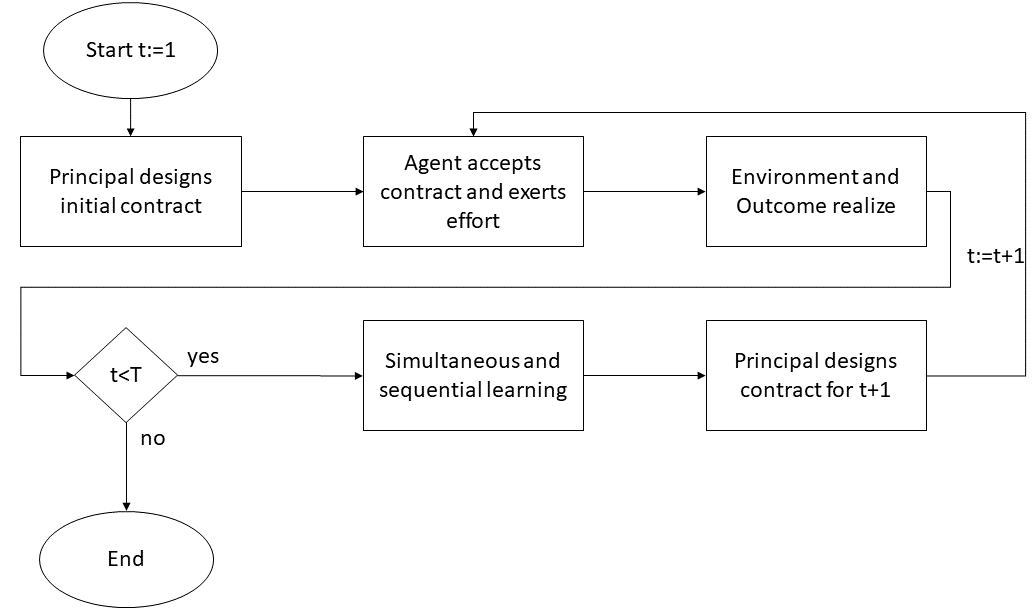}
	\caption{Flow diagram }
	\label{fig:flow}
\end{figure}

We transfer the standard hidden-action model \cite{Holmstrom.1979} into an agent-based model following the agentization procedure introduced by Guerrero and Axtell \cite{Guerrero.2011} and Leitner and Behrens \cite{Leitner.2015}. We limit the principal's and the agent's availability of information about the environment and endow them with the capability to learn about the environment over time. As a consequence of this change, the principal and the agent can no longer find the optimal (second-best) contract immediately but search for it over time. This stepwise search for the optimal contract requires us to switch from a one-periodic to a multi-periodic model in which the principal and the agent agree upon a contract in every timestep. Due to the employed learning mechanism (which is introduced below), the principal's and the agent's states of information are changing over time. We do not give the agent the possibility to allocate effort over multiple periods. A condensed overview of the sequence of the events is provided in Figure \ref{fig:flow}.

We indicate time steps by $ t={1,...,T} $. After the adjustment to a multi-periodic model, the principal's utility function introduced in (\ref{utilityP}) takes the form of

\begin{equation}
\label{utilityP-abm}
U_P(x_t,s(x_t))=x_t-s(x_t)~,
\end{equation}

\noindent where $ x_t $ denotes the outcome and $ s(x_t) $ is the agent's compensation in time step $ t $. The production function introduced in (\ref{prodctionfunction}) is adapted, so that

\begin{equation}
\label{outvome_abm}
x_t=a_t+\theta_t~,
\end{equation}

\noindent where $ a_t $ stands for the effort-level selected by the agent from a set of all feasible actions $ A_t $ in $ t $. $ A_t $ is a subset of $ A $ with the range from zero to the value $ a_t $, whereby the latter refers to the effort-level which is required to achieve the second-best solution according to the standard hidden-action model. Setting the boundaries in this way is in line with \cite{Holmstrom.1979} and assures the feasibility of the solution. We denote the effort level for which the principal designs the contract by $\tilde{a}_t$ and refer to it as incited effort. The variable $ \theta_t $ denotes the realized exogenous factor in $ t $ which follows a Normal distribution. Given these adaptations, the agent's compensation function $ s(x_t) $ takes the form of 
\begin{equation}
s(x_t)= x_t*p_t~,
\end{equation}
where $ p_t \in [0,1] $ is the premium parameter in \(t\). After the adaption to a multi-periodic model, the agent's utility function takes the form of
\begin{equation}
U_A(s(x_t),a_t)= \overbrace{\dfrac{1-e^{-\eta*s(x_t)}}{\eta}}^{V(s(x_t))}-\overbrace{\dfrac{a_t^2}{2}}^{G(a_t)}~,
\end{equation}
where $ \eta  $ represents the agent's Arrow-Pratt measure of risk-aversion \cite{Arrow.1964}. The agent strives for maximizing this utility function but, due to the adaptions outlined below, replaces the environmental factor $\theta_t$ by his expectation thereof. For the computation of the agent's expectation about the environmental factor see  (\ref{calcThetaAgent}) and (\ref{theta_abmA}). 

The most central change during the process of agentization is the relaxation of the availability of information about the environment for the principal and the agent. In the standard hidden-action model \cite{Holmstrom.1979}, it is assumed that both parties have knowledge about the distribution of the exogenous factor and its parameterization, so that they can compute an expected value for the environmental variable. In contrast to this, in the agent-based model variant they have to learn about the environment over time. The assumptions related to the environment are relaxed in the following way:
\begin{enumerate}
	\item The principal and the agent no longer have full information about the environmental factor at the beginning of the simulation.
	\item The principal and the agent are able to individually learn about the environment over time.
	\item The principal and the agent also have the cognitive ability to store the gathered information about the environment in their memory.
\end{enumerate}
These changes are implemented by a simultaneous and sequential learning model (see also \cite{Leitner.2020}). Part of this learning model is not only the process of gathering and storing the information but also the application of this information to compute the estimation of the exogenous factor. Both the principal and the agent are now able to individually learn about the environment and estimate the exogenous factor in the same way: They make their conclusions about the environment on the basis of the observable outcome $ x_t $ (see (\ref{outvome_abm})), which is possible because the only unknown information in the equation is the realized environmental factor. Recall, both the principal and the agent know the realized outcome $ x_t $, the agent knows the exerted effort $ a_t $, and the principal knows the incited effort $ \tilde{a}_t  $. Using this information, the principal and the agent can infer the estimations of the environmental factor from the realized outcome, according to 
\begin{equation}
\label{calcThetaPrincipal}
\tilde{\theta_t}=x_t- \tilde{a}_t ~,
\end{equation}
\noindent and

\begin{equation}
\label{calcThetaAgent}
\theta_t= x_t-a_t~,
\end{equation}
\noindent respectively. Furthermore, they are able to privately store the estimations in their memory until their defined cognitive capacity ($ m_P $ for the principal and $ m_A $ for the agent) is reached. Once the cognitive capacity is reached, the oldest entries are replaced by the most recent observations. Their expectation about the environmental factor is computed by averaging all privately stored and retrievable estimations of exogenous factors, so that
\begin{equation}
\hat{\theta}_{Pt}=
\begin{cases}
\frac{1}{t-1}~ \sum\limits_{n=1}^{n=t-1}{\tilde{\theta}}_{n} &\text{if~} m_P=\infty,\\
\frac{1}{m_P} \sum\limits_{\substack{\forall t \leq m_P: n=1 \\ \forall t > m_P: n = t-m_P }}^{n=t-1}{\tilde{\theta}}_{n} &\text{if~} m_P<\infty~,\\
\end{cases} 
\label{theta_abmP}
\end{equation} 
\noindent for the principal and
\begin{equation}
\hat{\theta}_{At}=
\begin{cases}
\frac{1}{t-1}~ \sum\limits_{n=1}^{n=t-1}{\theta}_{n} &\text{if~} m_A=\infty,\\
\frac{1}{m_A} \sum\limits_{\substack{\forall t \leq m_A: n=1 \\ \forall t > m_A: n = t-m_A }}^{n=t-1}{\theta}_{n} &\text{if~} m_A<\infty~,\\
\end{cases} 
\label{theta_abmA}
\end{equation} 
for the agent.

Next, the principal randomly discovers two alternative effort levels in the search space $ A_t $ which together with $\tilde{a}_t$ serve as candidates for $\tilde{a}_{t+1}$. Using her expectation about the environment $\hat{\theta}_{Pt}$, the principal computes the expected outcome $\tilde{x}_{Pt}$ (according to  (\ref{outvome_abm})) and her expected utility (according to (\ref{utilityP-abm})) associated with all three candidates.  Notice that for all effort levels the associated premium parameter is computed according to 
\begin{equation}
p_t = \max_{p=[0,1]} U_p(\tilde{x}_{Pt},s(\tilde{x}_{Pt}))~.
\end{equation}
Finally, the principal selects the candidate with the highest expected utility as desired effort level $ \tilde{a}_{t+1} $ for period $ t+1 $ and communicates the associated premium parameter $ p_{t+1} $  to the agent. In period $ t+1 $, the agent starts over the procedure with selecting an effort level $ a_{t+1} $ using $ p_{t+1} $ (see Figure \ref{fig:flow}).

\begin{table}
	\caption{Notation for the agent-based model variant}
	\label{tab:abm}       
	\begin{tabular}{lll}
		\hline\noalign{\smallskip}
		Description& Parameter \\
		\noalign{\smallskip}\hline\noalign{\smallskip}
		Timesteps & $ t $ \\
		Principal's utility & $ U_P $ \\
		Agent's utility & $ U_A $ \\
		Agent's Arrow-Pratt measure of risk-aversion & $ \eta $ \\
		Agent's share of outcome in $t$ & $ s(x_t)=x_t*p_t $ \\
		Outcome & $x_t=a_t+\theta_t$ \\
		Principal's expected outcome & $ \tilde{x}_{Pt} $\\
		Premium parameter in $t$ & $p_t$\\
		Exerted effort level in $t$ & $ a_t $\\
		Induced effort level by the principal in $t$ & $ \tilde{a}_t $\\
		Set of all feasible actions in $t$ & $ A_t $ \\
		Exogenous (environment) variable in $t$ & $\theta_t$\\
		Principal's estimation of the realized exogenous factor in $t$ & $\tilde{\theta_t}$ \\
		Mental capability of the principal & $m_P$\\
		Mental capability of the agent & $m_A$\\
		Averaged expected exogenous factor of the principal & $ \hat{\theta}_{Pt} $\\
		Averaged expected exogenous factor of the agent & $ \hat{\theta}_{At} $\\
		
		\noalign{\smallskip}\hline
	\end{tabular}
\end{table}
\section{Simulations parameters and Results}
\label{sec3}
We have conducted 8 scenarios with different underlying assumptions about the cognitive capacity (memory) of the principal and the agent and two different levels of environmental turbulence.

Recall that the environmental variable is modeled to follow a Normal distribution. The variations in environmental turbulence are operationalized by altering this distribution's standard deviation which is set relative to the optimal outcome $ x $ of the standard hidden-action model (second-best solution in \cite{Holmstrom.1979}).
For the case of a rather stable environment we set  $ \sigma = 0.05x $ and for a rather unstable environment we set $ \sigma=0.45x $. The distribution's mean is fixed at zero. 

In the 8 investigated scenarios, we have, in general, two configurations related to the cognitive capacity of the principal and the agent. The first one assumes that the principal has advantage in cognitive capacity and the second assumes the opposite situation. Each of them consist of 4 scenarios where the one in advantage always has unlimited memory and the other one has either a memory with the length of one or five. The agent's Arrow-Pratt measure is always set to $ \eta=0.5 $, which characterizes a risk-averse agent. 
 
 For every scenario, we perform 700 repetitions ($ R=700 $), our analysis focuses on the first 20 time steps ($ t=1,...,20 $). In order to implement our simulation model we use MATLAB\textsuperscript{\textregistered}. 

We report the averaged normalized effort level carried out by the agent in every period $ t $ as performance measure. Therefore in every simulation run $ r=1,...,R $ and for every period $ t=1,...,T $ we track the effort level $ a_{tr} $ chosen by the agent, and normalize it by the optimal level of effort $ a* $.
\begin{equation}
\label{eq:performanceIndi}
\phi_t = \dfrac{1}{R}\sum_{r=1}^{r=R}\dfrac{a_{tr}}{a*}
\end{equation}
We report the averaged normalized effort level as a measure of performance, as provides fundamental insights into the functioning of the emergent incentive schemes without further perturbation caused by environmental turbulence.

\subsection{Results 1: Advantage in information for the agent}
\begin{figure}[h]
	\centering
	\includegraphics[scale=0.4]{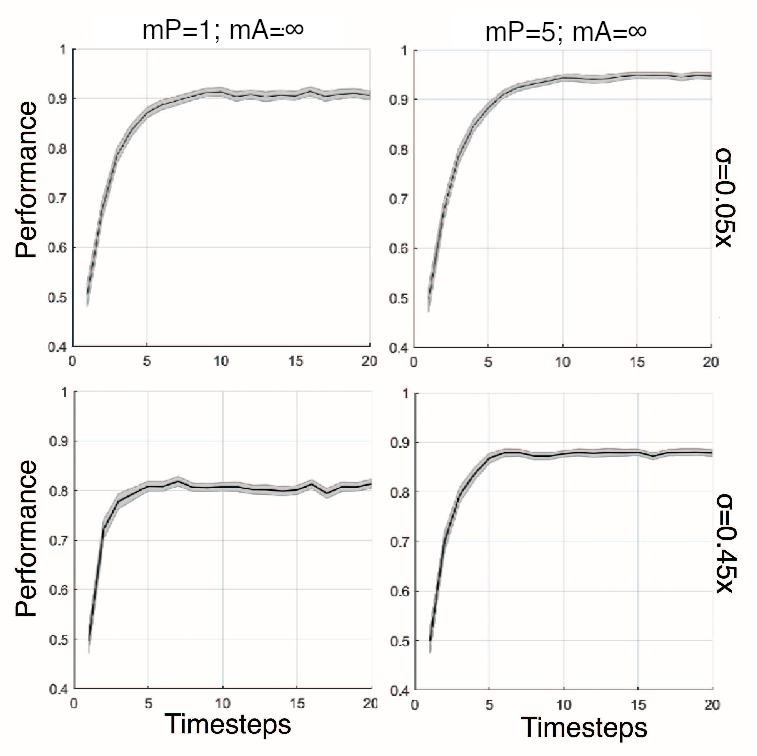}
	\caption{Situations with advantage regarding cognitive capacity for the agent (A) and the cognitive capacity of the principal (P) with either $ m_P=1 $ or $ m_P=5 $. Scenarios a plotted for two different environmental situations, stable ($ \sigma=0.05x $) and unstable $ \sigma=0.45x $..}
	\label{fig:aunlp15}
\end{figure}

First, we analyze the results in the situation with the advantage for the agent. In the two scenarios with a relatively stable environment ($\sigma=0.05x$) we can see that performance is significantly higher (at the end of the observation period) for cases in which the principal's cognitive capacity is higher, too (in terms of an increase in memory length): As the principal's memory increases from \(m_P=1\) to \(m_P=5\), the observed final averaged normalized effort level increases from $ 0.9063 $ to $ 0.9474 $ (see the two top subplots in Figure \ref{fig:aunlp15}). 
We can observe similar results for the scenarios with a rather unstable environment ($\sigma=0.45x$). Here, the final performance measure increases from $ 0.8135 $ to $ 0.8785 $ as the principal's memory increases from \(m_P=1\) to \(m_P=5\) (see the two bottom subplots in Figure \ref{fig:aunlp15}). Furthermore, in line with intuition, we can observe that an increase in environmental turbulence leads to a decrease in the effort exerted by the agent.

We further analyze the effect of the principal's memory on the average variance of the results (i.e., the average variance of the effort exerted by the agent throughout the entire observation period): For the scenarios in a rather unstable environment, the average standard deviation significantly decreases from $0.17405$ (for  $m_P=1$) to $0.15719$ (for $m_P=5$). For the cases in a rather stable environment no significant differences can be observed. The significance at the \(99\%\)-level was confirmed using an F-test. Thus, increasing the principal's cognitive capacity in an unstable environment significantly reduces the variance of the effort induced by the incentive scheme (and exerted by the agent). In other words, if the principal manages to increase her memory in an turbulent environment, she not only increases the average normalized effort level but also significantly reduces the risk of extreme deviations from this value.

Finally, we take a look on the stability of the averaged normalized effort level. We regard a solution to be stable as soon as (i) the averaged normalized effort level at period $ t $ is not significantly different from the same measure in period $ t-1 $, and given that (ii) this condition does not change after this point in time. For this analysis, we perform a T-test and set $\alpha=0.01 $.
First, we focus on rather stable environments: For cases with a low (high) cognitive capacity for the principal of $ m_P=1 $ and $ m_P=5 $, we observe a stable solution for periods $ t=7 $ and $ t=9 $ onwards, respectively.
For unstable environments, we can observe that a stable solution emerges earlier in both scenarios. For $ m_P=1 $ ($ m_P=5 $) the solution becomes stable in period $ t=4 $ ($ t=6 $). This is a counter-intuitive result as one would expect that turbulence in the environment leads to instability in the emergent solution. 

	\subsection{Result 2: Advantage in information for the principal}
\begin{figure}[h]
	\centering
	\includegraphics[scale=0.4]{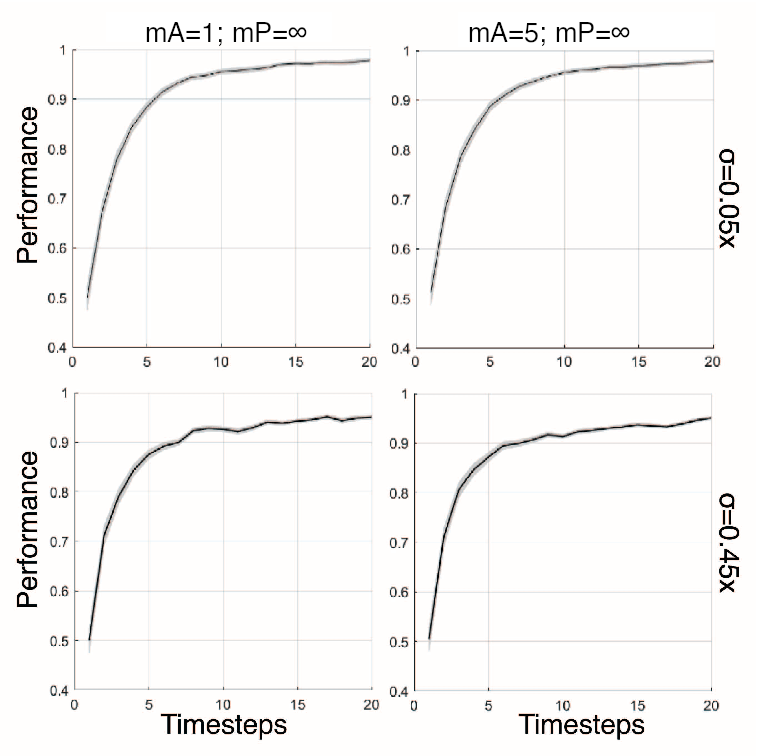}
	\caption{Scenarios with advantage regarding cognitive capacity for the principal (P) and the cognitive capacity of the agent (A) with either $ m_A=1 $ or $ m_A=5 $. Scenarios a plotted for two different environmental situations, stable ($ \sigma=0.05x $) and unstable $ \sigma=0.45x $. }
	\label{fig:punla15}
\end{figure}

This section analyses the cases in which the principal has an advantage in information. In stable environments ($ \sigma=0.05x $), we cannot observe significant differences: For the agent's memory of $ m_A=1 $ and $m_A=5$, the observed final averaged normalized effort levels are $ 0.9784 $ and $ 0.9786 $, respectively (see the two top subplots in Figure \ref{fig:punla15}). The same result  can be observed for unstable environments ($ \sigma=0.45x $). Here, the final performances are $ 0.9507 $ and $ 0.9512 $ for $ m_A=1 $ and $ m_A=5 $, respectively (see the two bottom subplots in Figure \ref{fig:punla15}). Additionally, in line with intuition, we can observe that an increase in environmental turbulence leads to a decrease in the effort exerted by the agent. 
 
 Further, we analyze the effects of the agent's memory on the average variance of the results (i.e., the average variance of the effort exerted by the agent throughout the entire observation period). For unstable environments ($ \sigma=0.45x $), the observed results are similar to the ones presented in the previous section: The variances of the exerted effort levels are significantly different at the 99\%-level for $ m_A=1 $ and $  m_A=5 $, for the former (latter) we observe a standard deviation of $ 0.1953 $  ($ 0.15781 $). For stable environmnents, the results indicate that increasing the agent's memory significantly decreases the standard deviation of the exerted effort at the 95\%-level: We observe $ 0.14089 $ and $ 0.13821 $ for $ m_A=1 $ and $  m_A=5 $, respectively. Thus, increasing the agent's cognitive capacity significantly reduces the variance of the effort induced by the incentive scheme (and exerted by the him). In other words, if the agent manages to increase his memory, he significantly reduces the risk of extreme deviations from the performance value.

Finally, we investigate the stability of the averaged normalized effort level. 
First, we focus on the scenarios with rather stable environments: For the cognitive capacity of the agent of $ m=1 $ and $ m=5 $ we reach the stable point at period $ t=14 $ and $ t=13 $, respectively. For unstable environments we can observe in both scenarios that a stable solution emerges earlier. For $ m=1 $ ($ m=5 $) the performance reaches a stable point at period $ t=9 $ ($ t=7 $). This leads to the same counter-intuitive result as in the section before, as one would expect that turbulence in the environment leads to instability in the emergent solution. 

\section{Conclusion and Future Work}
\label{sec4}

The results presented in this paper deliver some insights about the effects of heterogeneous agents in a hidden-action setting: 
\begin{itemize}
	\item Our results suggest, that gathering information about the environment is a good strategy for the principal to increase his utility especially situations in which the environment is rather turbulent.
	\item In turbulent environments, increasing the memory of both the principal and the agent always has a positive effect on the variance of the results. This means that increasing the memory significantly reduces the risk of extreme deviations from the performance measures reported above.
	\item  In stable environments, the results, on the contrary, suggest that only an increase of the agent's memory leads to an significant decrease in the exerted effort's variance.
	\item Surprisingly, the presented results indicate, that turbulence has a positive effect on stability, so that a stable solution emerges earlier in turbulent environments.
	
\end{itemize}

Future work might want to deeper investigate the effects of heterogeneous memory in the hidden-action setting (e.g, more memory length) and also include cognitive biases when characterizing the principal's and the agent's cognitive capabilities (such as the recency or the primacy effect \cite{Mantonakis.2009}). Another potentially fruitful option avenue for future research might be to limit the principal's knowledge about the characteristics of the agent (such as the utility function).

\section*{Acknowledgment}
This work was supported by funds of the Oesterreichische Nationalbank (Austrian Central Bank, Anniversary Fund, project number: 17930).
\bibliographystyle{IEEEtran}
\bibliography{bibio}

\begin{thebibliography}{10}
\providecommand{\url}[1]{#1}
\csname url@samestyle\endcsname
\providecommand{\newblock}{\relax}
\providecommand{\bibinfo}[2]{#2}
\providecommand{\BIBentrySTDinterwordspacing}{\spaceskip=0pt\relax}
\providecommand{\BIBentryALTinterwordstretchfactor}{4}
\providecommand{\BIBentryALTinterwordspacing}{\spaceskip=\fontdimen2\font plus
\BIBentryALTinterwordstretchfactor\fontdimen3\font minus
  \fontdimen4\font\relax}
\providecommand{\BIBforeignlanguage}[2]{{%
\expandafter\ifx\csname l@#1\endcsname\relax
\typeout{** WARNING: IEEEtran.bst: No hyphenation pattern has been}%
\typeout{** loaded for the language `#1'. Using the pattern for}%
\typeout{** the default language instead.}%
\else
\language=\csname l@#1\endcsname
\fi
#2}}
\providecommand{\BIBdecl}{\relax}
\BIBdecl

\bibitem{Holmstrom.1979}
B.~Holmstr\"om, ``{Moral hazard and observability},'' {The Bell Journal of
  Economics}, vol.~10, no.~1, 1979, p.~74.

\bibitem{Axtell.2007}
R.~L. Axtell, ``{What economic agents do: How cognition and interaction lead to
  emergence and complexity},'' {The Review of Austrian Economics}, vol.~20, no.
  2-3, 2007, pp. 105--122.

\bibitem{Schneeweiss.1987}
C.~Schneeweiss, ``{On a formalisation of the process of quantitative model
  building},'' {European Journal of Operational Research}, vol.~29, no.~1,
  1987, pp. 24--41.

\bibitem{Jensen.1983}
M.~C. Jensen, ``{Organization Theory and Methodology},'' {The Accounting
  Review}, vol.~58, no.~2, 1983, pp. 319--339.

\bibitem{Eisenhardt.1989}
K.~M. Eisenhardt, ``{Agency theory: An assessment and review},'' {The Academy
  of Management Review}, vol.~14, no.~1, 1989, p.~57.

\bibitem{Shapiro.2005}
S.~P. Shapiro, ``{Agency Theory},'' {Annual Review of Sociology}, vol.~31,
  no.~1, 2005, pp. 263--284.

\bibitem{Leitner.2020}
S.~Leitner and F.~Wall, ``{Decision-facilitating information in hidden-action
  setups: an agent-based approach},'' {Journal of Economic Interaction and
  Coordination}, 2020, pp. 1--36.

\bibitem{Guerrero.2011}
O.~A. Guerrero and R.~L. Axtell, ``{Using Agentization for Exploring Firm and
  Labor Dynamics},'' in {Emergent Results of Artificial Economics}, S.~Osinga,
  G.~J. Hofstede, and T.~Verwaart, Eds.\hskip 1em plus 0.5em minus 0.4em\relax
  Berlin, Heidelberg: {Springer Berlin Heidelberg}, 2011, pp. 139--150.

\bibitem{Leitner.2015}
S.~Leitner and D.~A. Behrens, ``{On the efficiency of hurdle rate-based
  coordination mechanisms},'' {Mathematical and Computer Modelling of Dynamical
  Systems}, vol.~21, no.~5, 2015, pp. 413--431.

\bibitem{Lambert.2001}
R.~A. Lambert, ``{Contracting theory and accounting},'' {Journal of Accounting
  and Economics}, vol.~32, no. 1-3, 2001, pp. 3--87.

\bibitem{Arrow.1964}
K.~J. Arrow, ``{The Role of Securities in the Optimal Allocation of
  Risk-bearing},'' {The Review of Economic Studies}, vol.~31, no.~2, 1964,
  p.~91.

\bibitem{Mantonakis.2009}
A.~Mantonakis, P.~Rodero, I.~Lesschaeve, and R.~Hastie, ``{Order in choice:
  effects of serial position on preferences},'' {Psychological science},
  vol.~20, no.~11, 2009, pp. 1309--1312.

\end{thebibliography}

\end{document}